\documentclass{article}
\usepackage{graphicx}
\usepackage{amsfonts}

\title{Symmetry-driven layered dynamics in the Kuramoto–Sivashinsky equation}

\author{Alessandro Barone \\ 
\\
        Department of Physics and Astronomy, \\University of Bologna, Bologna, Italy}

\begin{document}

\maketitle

\begin{abstract}
In this work, we uncover a layered organization of the state space in the Kuramoto–Sivashinsky equation with periodic boundary conditions, in which multiple invariant sets coexist at fixed system parameters and are selected by the initial condition. Within this framework, both chaotic attractors and periodic orbits (traveling waves) can be systematically generated by amplifying a single initial condition and parameterized by the initial energy. As the energy increases, the period of the periodic orbits decreases according to an inverse scaling law. In transitional parameter regions, periodic dynamics at low initial energy is found to coexist with strange attractors at higher energy levels, revealing a unique layered landscape governed by the viscosity and the initial condition. We conjecture that this behavior is linked to continuous spatial translational symmetry, which is reflected in the degeneracy of the neutral part of the Lyapunov spectrum.
\end{abstract}

\section{Introduction}

Symmetries have long played a central role in the scientific description of natural phenomena. In physics, mathematics and natural science, symmetries often reveal profound structural properties of the governing laws. A paradigmatic example is provided by mechanical systems, where invariance under time translations is a reflection of the conservation of energy, while spatial translational and rotational symmetries lead to the conservation of linear and angular momentum \cite{LandauConservationLaws}.

In dynamical systems, symmetries exert an equally profound influence on the qualitative behavior of the dynamics. Invariance under a group of symmetries can deeply influence the structure of the state space, the types of solutions that emerge, and the associated bifurcation scenarios \cite{syemmtries}. Moreover, continuous symmetries directly influence the tangent space structure by generating neutral directions in the linearized dynamics, associated with zero Lyapunov exponents (LEs) in the Lyapunov spectrum \cite{Pikovsky_Politi_2016}.

In this work we investigate the interplay between symmetry and multistability in the Kuramoto--Sivashinsky equation (KSE), a celebrated and well known nonlinear partial differential equation (PDE), given by: 
\begin{equation}
    u_t + u_{xx} + \nu u_{xxxx} + u u_x = 0 ,
    \label{eq:KS1}
\end{equation}
with periodic boundary conditions, $u(x,t) = u(x+L,t)$.

Kuramoto \cite{Kuramoto1976PersistentPO} introduced the KSE in the study of angular-phase turbulence in reaction--diffusion systems, while Sivashinsky \cite{SIVASHINSKY19771177} derived an equivalent equation in the context of weakly unstable laminar flame fronts. The KSE is an extensively studied model in mathematical physics, celebrated for the wide range of dynamical behaviors, including periodic, chaotic~\cite{papageorgiou_route_1991}, and intermittent regimes~\cite{Barone2025}. The structure of steady and traveling-wave solutions of the KSE has been systematically investigated since the seminal work of Greene and Kim~\cite{GreeneKim1988}.

In this system, both the size $L$ and the viscosity parameter $\nu \in \mathbb{R}^+$ act as control parameters. Throughout this work we fix $L = 10\pi$, so that $x \in [0,L]$, and investigate the system’s behavior as $\nu$ is varied.
The KSE possesses a rich symmetry structure, which is summarized in detail in \cite{KSCvi}. Among these symmetries, two are continuous and have important consequences for both the dynamics and the structure of the tangent space.
First, the time-translation invariance, typical of continuous autonomous systems, leads to the existence of a neutral direction tangent to the flow, associated to one zero LE \cite{Pikovsky_Politi_2016}.

Second, the periodic spatial domain induces a continuous symmetry under spatial translations, $u(x,t) \rightarrow u(x+\ell,t), \, \, \, \,  \ell \in [0,L)$, which generates an additional zero LE. 

The periodic boundary promotes a Fourier representation of the field $u(x,t)$ \cite{KSCvi}. Denoting by $\hat{u}_k(t)$ the Fourier coefficients of the solution and by $q_k = 2\pi k / L$ the corresponding wavenumbers, the KSE can be written in Fourier space as a system of coupled ordinary differential equations,
\begin{equation}
    \frac{\mathrm{d}\hat{u}_k}{\mathrm{dt}} =
    \left(q_k^2 - q_k^4\right)\hat{u}_k
    - \frac{i q_k}{2}
    \mathcal{F}_N \left[
        \left(\mathcal{F}_N^{-1}[\hat{u}]\right)^2
    \right]_k, 
\end{equation}
with the corresponding equation for the tangent dynamics: 
\begin{equation}
    \frac{\mathrm{d}\hat{w}_k}{\mathrm{dt}} =
    \left(q_k^2 - q_k^4\right)\hat{w}_k
    - i q_k
    \mathcal{F}_N \left[
        \mathcal{F}_N^{-1}[\hat{u}]
        \otimes
        \mathcal{F}_N^{-1}[\hat{w}]
    \right]_k.
\end{equation}

The system is integrated numerically using the fourth-order exponential time-differencing Runge--Kutta method (ETDRK4) \cite{KSCvi, kassam_fourth-order_2005, COX2002430}. Time integration is performed with a fixed time step, $\Delta t = 0.1$, the spatial domain is discretized using $n = 101$ grid points and all the simulations have a length of $10^5$ time steps.

\section{Lyapunov exponents and coexistence of chaotic attractors}

\begin{figure}[h!]
    \centering
    \includegraphics[width=0.5\textwidth]{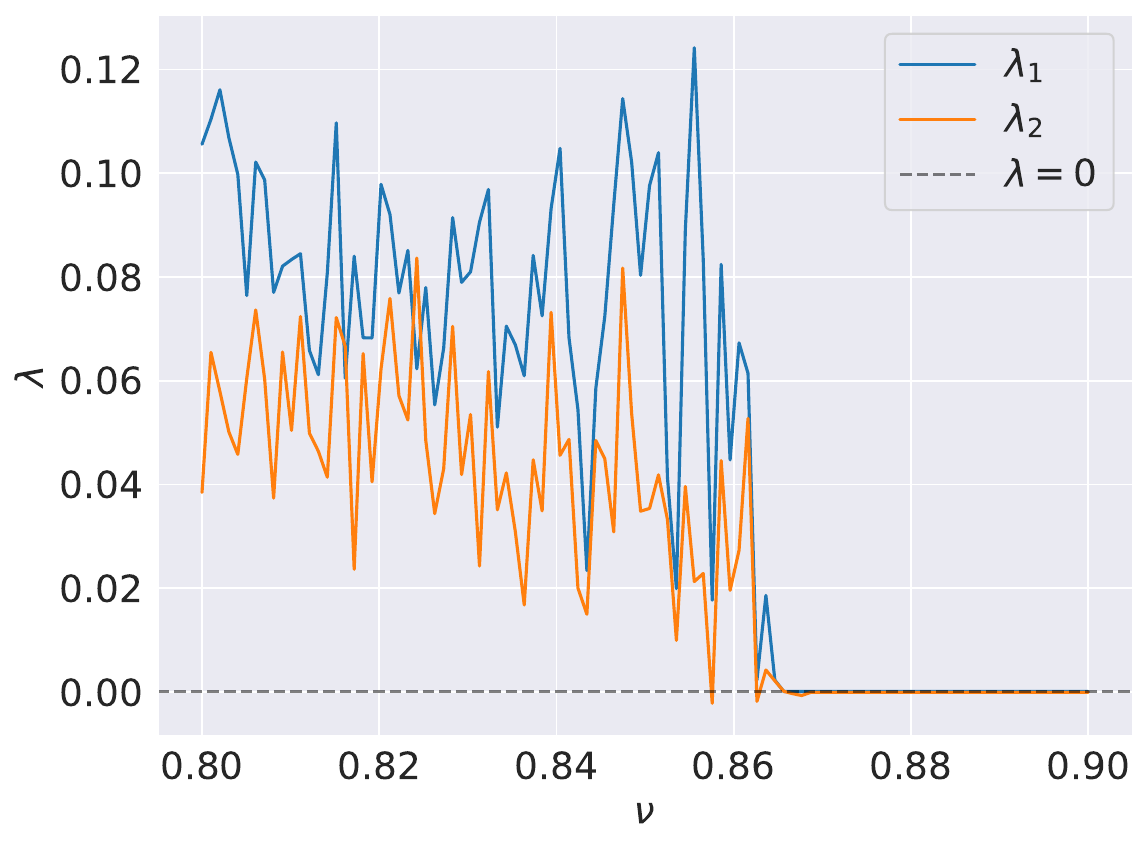}     % includes figure foo.eps
    \caption{First two Lyapunov exponents, $\lambda_1$ and $\lambda_2$, as a function of the viscosity parameter $\nu$. }
    \label{fig:le}
\end{figure}

An original motivational driver for pursuing this study arose from an unexpected behavior in the bifurcation diagram (not shown). We found that varying the control parameter $\nu$ did not reveal any recognizable structure: instead of organized branches, the diagram consisted of scattered points with no apparent pattern, even for parameter values (and initial conditions) for which the system was observed to converge to periodic solutions in the state space. This observation suggested that the system might not be fully described by a standard bifurcation scenario in $\nu$, as qualitative changes in the state space structure were also induced by variations in the initial condition. Early numerical investigations of the KSE already reported the coexistence of multiple solution branches and a strong sensitivity to initial conditions in symmetry-invariant settings~\cite{Kevrekidis1990}. 
Further evidence supporting this interpretation came from the pronounced sensitivity of the Lyapunov spectra on the viscosity parameter $\nu$ \cite{Barone2025}. This indicated that different initial conditions were not converging to the same attractor. Rather, the system was repeatedly selecting distinct invariant sets, even at fixed parameter values. Despite this sensitivity, the neutral part of the spectrum systematically displays a degeneracy, with two Lyapunov exponents equal to zero across all viscosity values considered, even for viscosity values where the system is observed to settle into periodic orbits ($\nu \ge 0.87$) (Fig.~\ref{fig:le}).

The main focus of this work is to shed light on a peculiar and yet unexplored dynamical feature. By varying the magnitude of the initial state vector while keeping the system parameters fixed at $\nu = 0.85$, the system converges to distinct chaotic attractors. In practice, the initial conditions act as an effective additional parameter. To illustrate this behavior, let us to consider a set of $20$ initial conditions obtained by systematically amplifying a reference initial state. Specifically, we define a family of initial conditions $u_0^{(k)} = \eta_k  U_0, \, \, \, \, \,  \eta_k = 10k, \quad k = 1,2,\ldots,20$ where $U_0$ is a reference initial condition sampled from the uniform distribution in the interval $[0,0.1]$. By computing the energy as the $L^2$ norm \cite{papageorgiou_route_1991}:
\begin{equation}
    \varepsilon(t) = ||u(x,t)||_2 = \sqrt{\int_0^Lu^2(x,t){\rm d}x},
\end{equation}

each initial condition can be characterized by a different initial energy, defined as $\varepsilon_0^{(k)} = ||u_0^{(k)}||_2$.

\begin{figure}[h!]
    \centering
    \includegraphics[width=0.7\textwidth]{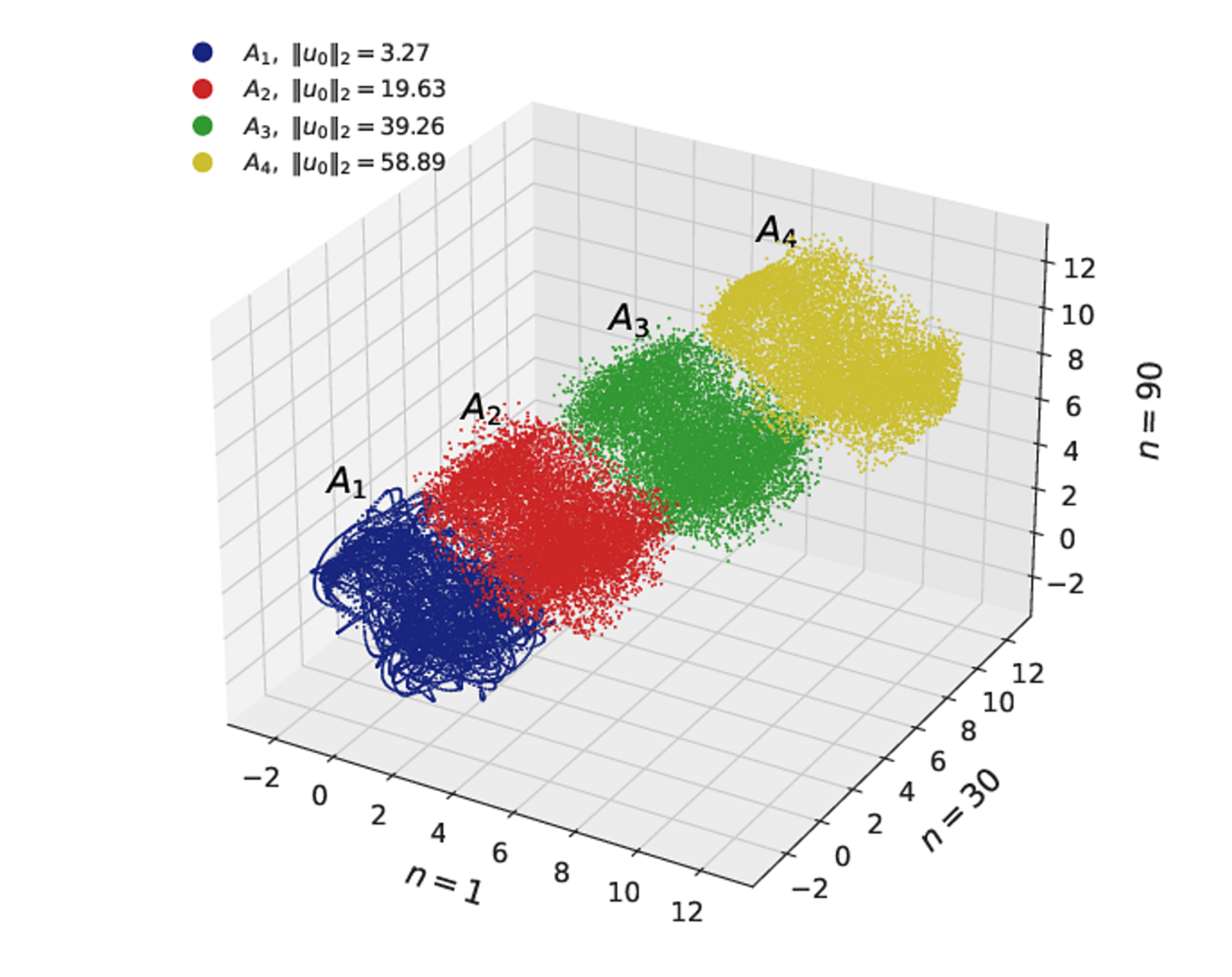} 
    \caption{Four chaotic attractors chosen out of 20 generated by the amplification of a single initial condition. On the top left the magnitude of the initial energy for each different attractor.}
    \label{fig:CHA1}
\end{figure}

In Fig.~\ref{fig:CHA1} we show four coexisting attractors obtained from four different initial conditions and for $\nu=0.85$, each characterized by a distinct initial energy. These examples are presented for illustrative purposes; however, by varying the initial condition through the amplification procedure discussed above, many additional attractors can be generated.

In our previous work we focused on identifying regions of the parameter space in which the KSE exhibits intermittent behavior (see Fig.~10 of Ref.~\cite{Barone2025}). In particular, intermittency was observed in the transition between periodic (traveling waves) and chaotic regimes. Within this parameter window, we further found that for values of the viscosity parameter close to the bifurcation point ($\nu \approx 0.865$), the KSE supports both periodic orbits, quasi-periodic orbits or intermittent state depending on the initial conditions. This gives rise to a rich dynamical landscape, in which the magnitude of the initial condition and the viscosity parameter $\nu$ jointly organize a wide range of possible dynamical scenarios, resulting in a highly structured and layered state-space organization. 
We conjecture that this behavior is linked to the continuous spatial translational symmetry. In fact, when the translational symmetry is broken by the adoption of non-periodic boundary conditions, the emergence of initial-condition-dependent orbits ceased. This suggests that it is the spatial invariance that provides the structural setting in which this phenomenon can occur and develop. The second zero Lyapunov exponent observed in the regimes considered here reflects precisely the presence of this additional neutral direction induced by spatial translational invariance, and it is not present in numerical setups where the BCs are not periodic.

%%#############################################################
\section{Periodic orbits}

For $\nu = 0.87$, the system exhibits a solution that appears as a closed trajectory in the full state space. 
In physical space, this solution corresponds to a traveling wave propagating along the spatial domain with constant velocity $c$~\cite{KSCvi}. 
Accordingly, in a reference frame moving at velocity $c$, the dynamics reduces to a fixed point, i.e., a relative equilibrium.

\begin{figure}[h!]
    \centering
    \includegraphics[width=0.5\textwidth]{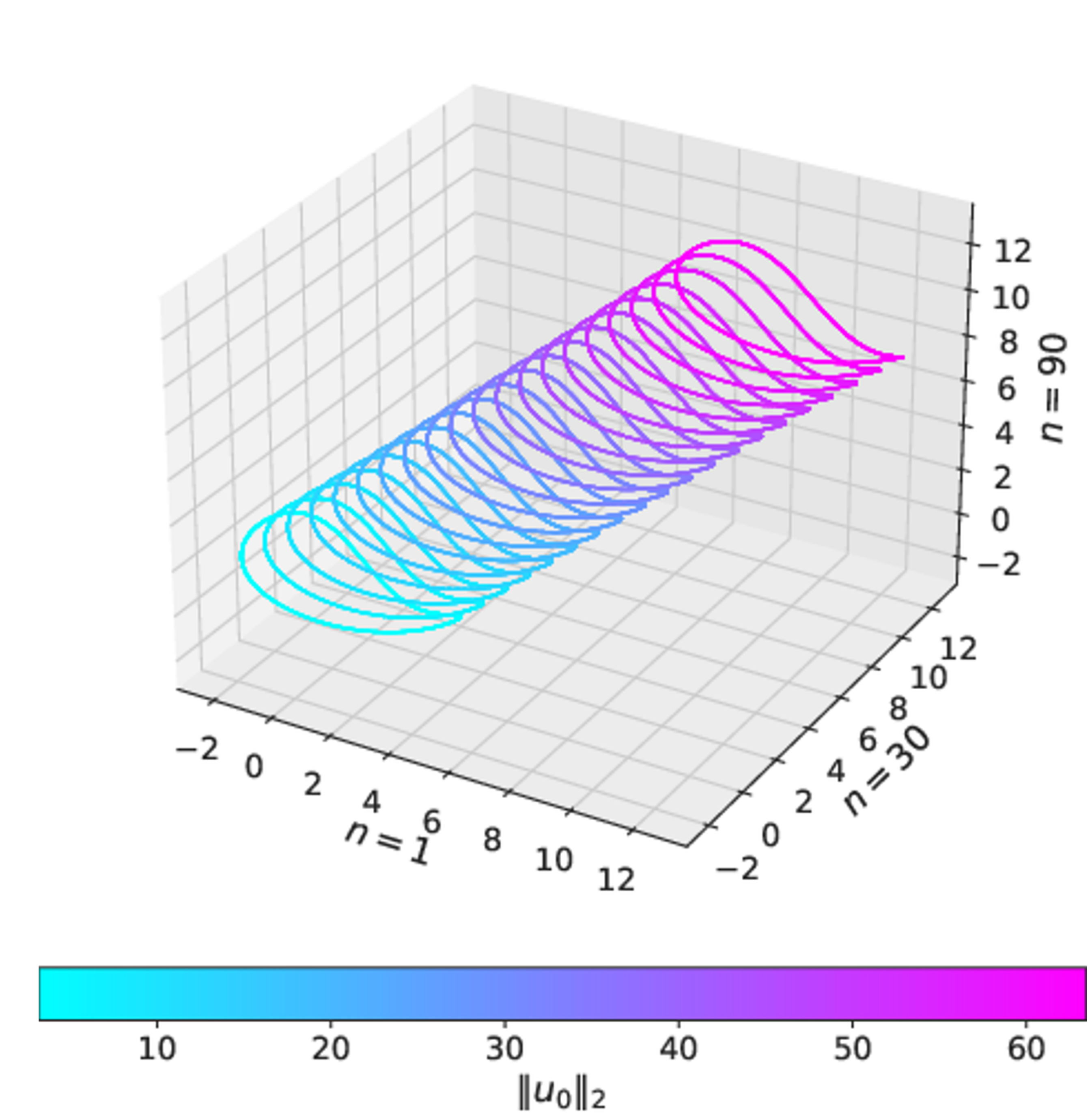}     % includes figure foo.eps
    \caption{Periodic orbits generated by $20$ different initial conditions (see text for details). Colors are proportional to the initial energy $\varepsilon_0^{(k)} = ||u_0^{(k)}||_2$ injected into the system.}
    \label{fig:orbits}
\end{figure}

In this scenario, we perform the amplification of a single initial condition, as explained above, obtaining $20$ independent simulations that converge to $20$ distinct periodic solutions, each uniquely associated with a specific value of the initial energy (Fig.~\ref{fig:orbits}). 
Notably, this procedure can be repeated for progressively smaller amplification factors of the initial condition, showing that the orbits are organized into a dense, tube-like structure in state space (Fig.~\ref{fig:orbits}), along which the magnitude of the initial condition acts as a parameter selecting the invariant set onto which the trajectory is confined.

\begin{figure}[h!]
    \centering
    \includegraphics[width=0.9\textwidth]{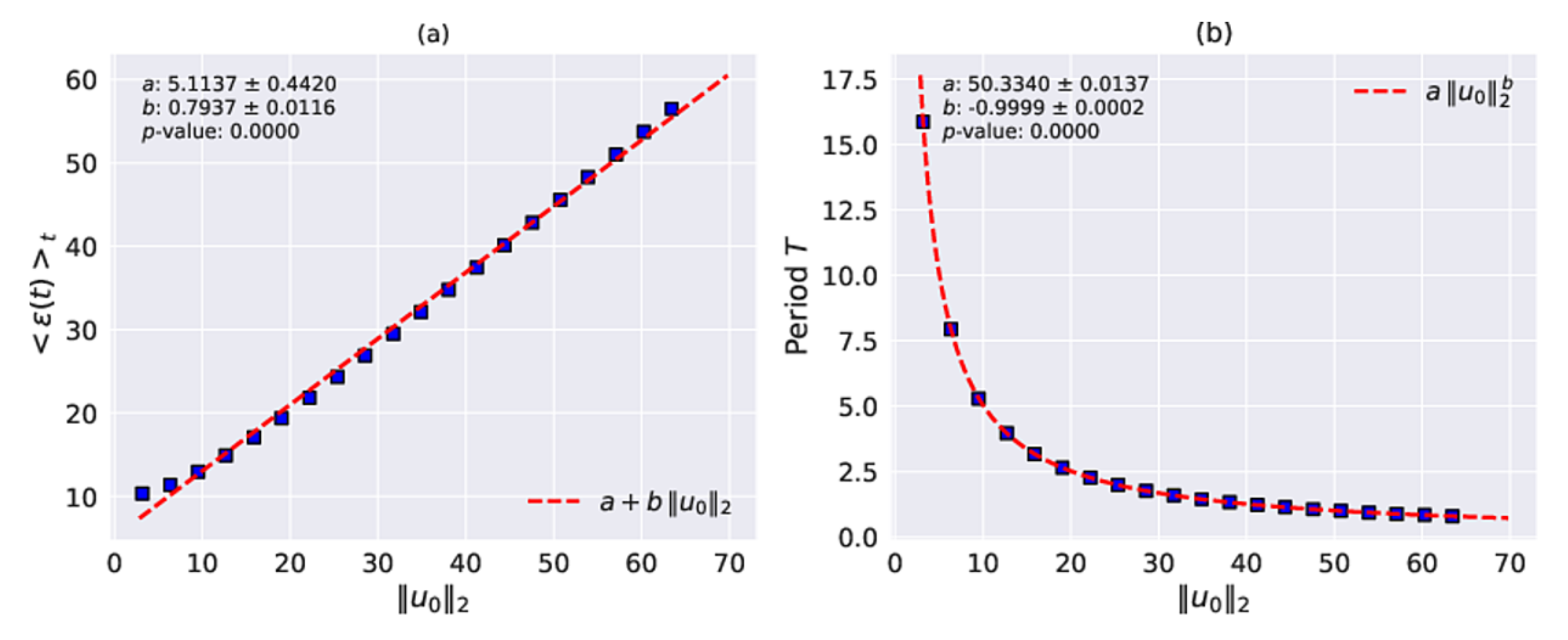}     % includes figure foo.eps
    \caption{Scaling of (a) the mean final energy of the system computed on a window of $\Delta t = 5  \times 10^3$ time steps at the end of the simulation with the initial energy and (b) the period of the periodic orbits with the initial energies.}
    \label{fig:scaling}
\end{figure}

Let us study how and if the periodic solutions change when the initial energy increases. Notably, they appear to vary in a systematic manner: the orbit amplitude, measured through the time-averaged energy computed on a window of $\Delta t = 5 \times 10^3$ time steps, increases approximately linearly with the initial energy (Fig.~\ref{fig:scaling}(a)), while the period decreases following an inverse scaling with the initial energy, $T \propto 1/||u_0||_2$ (Fig.~\ref{fig:scaling}(b)).

Finally, we performed an additional numerical experiment (not shown) in which a fixed initial condition $U_0$ is spatially reshuffled ten times. Although the resulting solutions remain very close in state space, they are nevertheless topologically distinct. This indicates that, while the period of the resulting dynamics scales with the initial energy, this information alone is not sufficient to uniquely determine the slice of the tube structure. Moreover, changing the functional shape of the initial condition, for instance to a cosinusoidal profile, leads to a qualitatively different asymptotic behavior, confirming that the selection of the solutions depends not only on the initial energy level but also on the function shape of the ICs.

\section*{Conclusions}

In this work we investigate the dynamics of the Kuramoto--Sivashinsky equation with periodic boundary conditions in regimes characterized by the presence of two zero Lyapunov exponents, associated with time-translation invariance and with the continuous spatial translational symmetry. 

Within this setting, we observe the coexistence of multiple invariant sets at fixed values of the control parameter ($\nu = 0.85$). By systematically amplifying a single reference initial condition, \emph{i.e.} by varying the magnitude of the initial state vector while keeping its shape fixed, the system converges to distinct chaotic
attractors. The same phenomenology is observed in parameter regions where the long-term dynamics is characterized by periodic orbits in state space, corresponding to traveling-wave solutions in physical space.

In these regimes, the corresponding invariant solutions are organized along a tube-like structure, that are parameterized by the initial energy measured by the squared $L^2$ norm. Along this family, we observe robust scaling behaviors: increasing the injected energy is associated with a decrease of the characteristic period, approximately following $T \sim \|u_0\|_2^{-1}$. Different functional shape of the ICs lead to distinct topological organizations in state space, revealing a layered structure arising from the combined role of the viscosity parameter and the initial condition.

\section*{Acknowledgments}
The author acknowledges Alberto Carrassi for his help and for reviewing this paper
prior to submission. This research was supported by the Scale-Aware Sea Ice Project
(SASIP), funded by Grant No.\ G-24-66154 of Schmidt Sciences, LLC.

\end{document}